\documentclass[aps,pre,twocolumn,showpacs]{revtex4}
\usepackage{epsfig}
\usepackage{times}
\begin{document}

\title{Decelerated invasion and waning moon patterns in public goods games with delayed distribution}

\author{Attila Szolnoki$^1$ and Matja{\v z} Perc$^2$}
\affiliation
{$^1$Institute of Technical Physics and Materials Science, Research Centre for Natural Sciences, Hungarian Academy of Sciences, P.O. Box 49, H-1525 Budapest, Hungary\\
$^2$Faculty of Natural Sciences and Mathematics, University of Maribor, Koro{\v s}ka cesta 160, SI-2000 Maribor, Slovenia}

\begin{abstract}
We study the evolution of cooperation in the spatial public goods game, focusing on the effects that are brought about by the delayed distribution of goods that accumulate in groups due to the continuous investments of cooperators. We find that intermediate delays enhance network reciprocity because of a decelerated invasion of defectors, who are unable to reap the same high short-term benefits as they do in the absence of delayed distribution. Long delays, however, introduce a risk because the large accumulated wealth might fall into the wrong hands. Indeed, as soon as the curvature of a cooperative cluster turns negative, the engulfed defectors can collect the heritage of many generations of cooperators, and by doing so start a waning moon pattern that nullifies the benefits of decelerated invasion. Accidental meeting points of growing cooperative clusters may also act as triggers for the waning moon effect, thus linking the success of cooperators with their propensity to fail in a rather bizarre way. Our results highlight that ``investing into the future'' is a good idea only if that future is sufficiently near and not likely to be burdened by inflation.
\end{abstract}

\maketitle

Large segments of our economy rely on continuous investments into a common pool that, at least so we are told, will or are very likely to pay off sometime in the future. Retirement plans and investments into the public health care and education system fall under this category, while somewhat shorter-term examples might include investing some fraction of personal wealth into a startup business or becoming a shareholder of a company with bright prospects. Either way, more often than not, the payoff from an investment is likely not going to be immediate, but rather it will set in with a delay.

The public goods game \cite{perc_jrsi13} captures the essential social dilemma that is linked with investments into a common pool. Those that contribute are cooperators, while those that do not are defectors. All the contributions within a group are multiplied to take into account synergetic effects of cooperation, and the resulting amount is divided equally among all group members irrespective of their strategies. If nobody invests, the group fails to harvest the benefits of a collective investment, and the society may evolve towards the ``tragedy of the commons'' \cite{hardin_g_s68}. Recently identified ways of avoiding this include the introduction of volunteering \cite{hauert_s02}, social diversity \cite{santos_n08}, bipartiteness \cite{gomez-gardenes_c11, gomez-gardenes_epl11}, random exploration of strategies \cite{traulsen_pnas09}, repeated group interactions \cite{van-segbroeck_prl12}, the risk of collective failures \cite{santos_pnas11}, selection pressure \cite{pinheiro_njp12}, conditional strategies \cite{szolnoki_pre12}, rare malicious agents \cite{arenas_jtb11}, as well as multiplexity \cite{wang_z_epl12, gomez-gardenes_srep12, gomez_prl13, wang_z_srep13}.

The game in its traditional form, however, does not take into account delays that might occur before the investments can actually be capitalized upon. Taking such delays into account essentially means altering the time scales of evolutionary dynamics \cite{pacheco_jtb06, roca_prl06, tanimoto_pre12}. Previous research showed that the diversity in reproduction time scales \cite{wu_zx_pre09b} and the coevolution of time scales \cite{rong_pre10} both promote cooperation. Indirect ways of altering the time scales, for example by means of breaking the symmetry between interaction and replacement \cite{ohtsuki_prl07}, wealth accumulation \cite{chadefaux_pone10}, consideration of long-term benefits \cite{brede_pone13}, or other coevolutionary processes \cite{wu_t_epl09, lee_s_prl11, perc_bs10}, have also been noted as beneficial for resolving social dilemmas. The application of a delayed distribution of accumulating goods, however, cannot be considered as a simple separation of time scales between learning (strategy adoption) and interaction (payoff accumulation) processes, because cooperators, as we will see, loose their payoff permanently between two consecutive distributions, while the effective payoff of defectors remains unchanged.

Here we therefore study the evolution of cooperation in the public goods game on the square lattice \cite{brandt_prsb03}, whereon initially each player on site $x$ is designated either as a cooperator ($s_i = C$) or defector ($s_i = D$) with equal probability. Regardless of their strategies, players receive a return on their investments from five overlapping groups of size $G=5$ \cite{szolnoki_pre09c}, but they do so only after each $\tau$-th round of the game. The payoff to each player stemming from a particular group is then $r\sum_{i,k} c_{i,k} /G$, where the sum runs over all the players within a group ($i=1 \ldots G$) and over $\tau$ rounds of the game ($k=1 \ldots \tau$), and $c_{i,k}$ is the contribution of player $i$ to the common pool during the $k$-th round. Without loss of generality cooperators contribute $1$ during each round of the game while defectors contribute nothing.

Unlike the distribution of public goods, the strategy transfers between randomly selected neighbors are possible during each round of the game, and they proceed according to the probability $w=(1+\exp[(P_i-P_j)/K])^{-1}$, where $K=0.5$ determines the uncertainty by strategy adoptions \cite{szolnoki_pre09c}. To take properly into account the seasonality that is introduced due to the delayed distribution of accumulating goods, the payoffs $P_i$ entering the Fermi function $w$ are averaged over the last $\tau$ rounds of the game. Note that this is necessary because the impact of each newly acquired payoff should last until the next distribution. During each full Monte Carlo step (MCS) all players will have a chance to pass their strategy once on average. We have used the square lattice having $L=200$ to $4800$ linear size and up to $10^6$ MCS before determining the fraction of cooperators $f_C$ within the whole population. Since the model is very simple, introducing only a single additional parameter $\tau$ to the game, the results should not be seen as describing a particular situation, but rather outline general consequences that may set in due to delayed distributions in the realm of collective social dilemmas.

\begin{figure}
\centerline{\epsfig{file=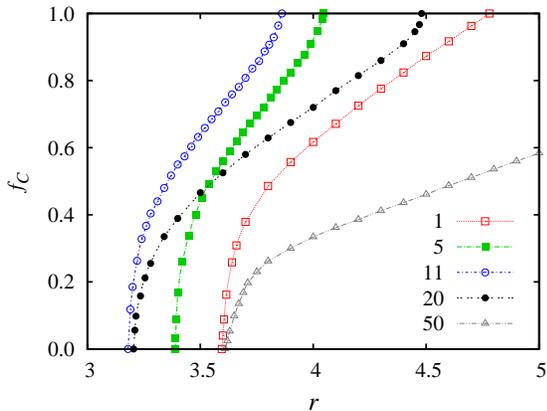,width=8cm}}
\caption{Fraction of cooperators $f_C$ in dependence on the multiplication factor $r$ for different values of the distribution delay $\tau$, as indicated in the figure legend. The positive trend for small and intermediate delays reverses when long delays are applied.}
\label{first}
\end{figure}

Figure~\ref{first} shows how delays in distribution affect the evolution of public cooperation. It can be observed that short and intermediate delays significantly lower the minimally required multiplication factor $r$ that is needed for cooperators to survive and to dominate in the population. By comparing the results for $\tau=11$ and $20$, however, we find that the trend reverses, affecting particularly strong the threshold for cooperator dominance. For $\tau=50$ cooperators are no longer able to eliminate defectors within the $r \leq G$ limit, while the threshold value of $r$ for the survival of cooperators raises back to the $r_c$ value recorded at $\tau=1$.

\begin{figure}
\centerline{\epsfig{file=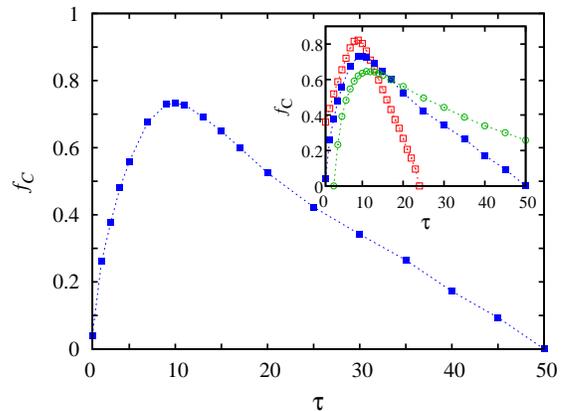,width=8cm}}
\caption{Fraction of cooperators $f_C$ in dependence on the distribution delay $\tau$ for $r=3.6$. An optimal value of $\tau$ provides the best conditions for the evolution of cooperation. Inset shows results obtained for $K=0.1$, $0.5$ and $0.8$, from left to right.}
\label{bell}
\end{figure}

The non-monotonous impact of delays in the distribution of public goods is depicted in Fig.~\ref{bell}, where the bell shaped outlay of the $f_C$ versus $\tau$ dependence indicates two competing effects affecting the evolution of cooperation. On the up side, delays help cooperators because they prevent defectors to utilize a temporarily high payoff via a fast invasion. Although cooperators are also negatively affected by the delays because they are unable to capitalize immediately on their investments, defectors are ultimately the ones who suffer more from the decelerated invasions capabilities. We emphasize that the invasion of cooperators is always slower because it needs a collective effort, and hence the slow down of the speed of invasions effectively enhances network reciprocity. Qualitatively similar observations were made in \cite{szolnoki_njp12}, where a less aggressive punishment strategy outperformed rewarding, in \cite{szolnoki_pre09}, where the reduced speed of invasion due to aging affected defectors more adversely than cooperators, as well as in \cite{wang_z_epl12}, where the suppressed feedback of individual success due to biased utility functions lead to a spontaneous separation of characteristic time scales of the evolutionary process on interdependent networks. From a different point of view, the described mechanism can also be interpreted as a ``memory effect'' (referring to the memory of preceding payoffs, not strategies), because a defector cannot utilize its potentially high current payoff due to preceding smaller payoffs, hence precluding a fast invasion. In general, cooperation is promoted because the aggressive invasion of defectors is more sensitive to the slowing-down than the build-up of collective efforts in sizable groups.

\begin{figure}
\centerline{\epsfig{file=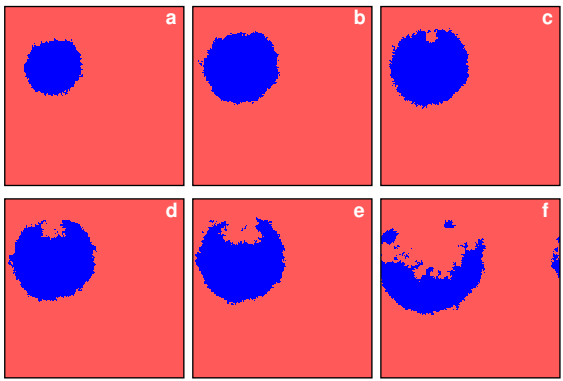,width=7.5cm}}
\caption{The waning moon effect, emerging spontaneously on a growing cooperative cluster (blue), as observed at $\tau=50$ and $r=3.65$. The cluster grows until its border has positive curvature [panels (a) and (b)]. As soon as the curvature becomes negative [panel (c)], which happens by chance due to the fluctuations along the expanding interface, defectors (red) can invade the bulk of the cluster. The cluster keeps growing on the other end, but the invasion of defectors is faster, resulting in the ``waning moon'' pattern. The total area occupied by cooperators decreases until somewhere again a small cooperative cluster with positive curvature around the entire border emerges (not shown). From that point onwards the whole cycle starts anew. Snapshots of the $200 \times 200$ square lattice in panels (b) -- (f) were taken $200$, $250$, $300$, $350$ and $600$ MCS after panel (a).}
\label{moon}
\end{figure}

While the up side of delays for the evolution of public cooperation is rooted in decelerated invasion, the down side manifests in a more subtle manner. Spatial patterns presented in Fig.~\ref{moon} provide vital insights. Even for low values of $r$, cooperative clusters are able to grow in the sea of defectors due to enhanced network reciprocity [panels (a) and (b)], as described above. Due to uncertainties in the strategy adoption, however, the front separating cooperators and defectors is not completely sharp, and it may happen that somewhere the curvature becomes negative. As panels (c) -- (f) illustrate, this becomes the entry point for defectors to invade. A characteristic waning moon pattern emerges, and accordingly we dub this the waning moon effect, which is intimately related with the delayed distribution of accumulating public goods. Namely, as soon as the curvature becomes negative, a lucky defector becomes surrounded by cooperators, and as such is scheduled to receive a huge payoff that is due to the preceding collaborative efforts. In fact, it is about to receive the heritage of many generations of cooperators, and this will have a positive impact on its payoff also in the following $\tau$ rounds of the game, despite the fact that the neighborhood will gradually turn defective. Ironically, the high payoff should have gone to a cooperator that contributed previously, but had turned defector before the distribution round. The lucky defector will also have an evolutionary advantage over neighboring cooperators, because the latter will continue to contribute to the common pool. Although due to enhanced network reciprocity cooperators may continue to spread at the other side of the moon, their propagation speed will be lower than the invasion of defectors, ultimately resulting in the waning moon pattern. The only chance for cooperators not to disappear entirely is if a small fraction of the previous homogeneous domain remains bounded by a front with positive curvature, at which point the whole process starts again. It is worth mentioning that the emergence of the waning moon pattern is a nice example of when the movement of a front separating two homogeneous phases depends sensitively on the sign of the curvature.

\begin{figure}
\centerline{\epsfig{file=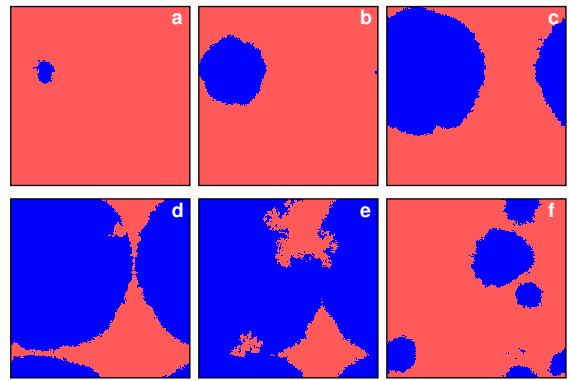,width=7.5cm}}
\caption{Growing cooperative clusters [panels (a), (b) and (c)] may collide, naturally giving rise to interfaces with negative curvature [panel (d)], and thus triggering the waning moon effect. Defectors are therefore able to invade [panel (e)], until small cooperative clusters re-emerge and start growing again [panel (f)]. The borders of these clusters will collide anew and so the process is repeated. Note that initially the cooperative cluster actually collides with itself due to periodic boundary conditions. We have intentionally applied a small system size for clarity, although of course the same collapse of independently growing domains can be observed for larger $L$ as well. Parameters are the same as in Fig.~\ref{moon}, and panels (b) -- (f) were taken $550$, $1250$, $1800$, $2020$ and $2630$ MCS after panel (a).}
\label{meet}
\end{figure}

Besides the spontaneous emergence of negative curvature due to fluctuations in the cooperative front, the latter also appears whenever two growing domains meet, as illustrated in Fig.~\ref{meet}. Due to the same reasons as described above, this will open the gate for the invasion of defectors [panels (d) and (e)], and again the only hope for cooperators is the emergence of smaller homogenous clusters with overall positive curvature [panel (f)]. Since the successful growth of cooperative clusters is thus inseparably linked with their failure in an intricate and quite bizarre turn of events involving large-scale spatial patterns, it is important to use sufficiently large system size $L$. This lowers the amplitude of fluctuations of $f_C$, firstly because then the growth and decline of cooperative cluster is unlikely to be synchronized in time, but also because more frequent collisions prevent the emergence of extreme, nearly homogeneous states. Regardless of what triggers the waning moon effect, it is straightforward to see that both the probability as well as the severity of such an event increase as we increase $\tau$. Accordingly, larger $\tau$ gradually drown out the positive effects stemming from enhanced network reciprocity by evoking the waning moon effect more frequently and in a more potent form, in turn giving rise to the bell-shaped dependence depicted in Fig.~\ref{bell}. Evidently, these phenomena are rooted in pattern formation on structured populations, and as such can not emerge in well-mixed systems.

\begin{figure}
\centerline{\epsfig{file=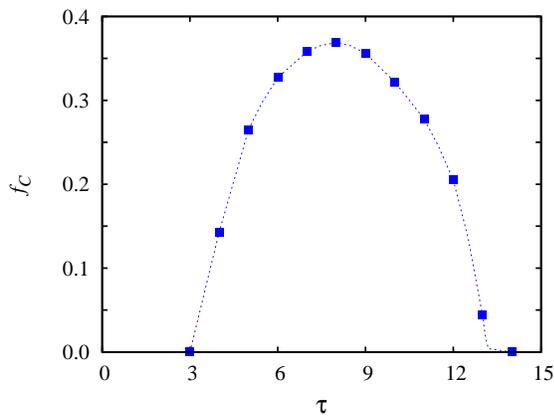,width=8cm}}
\caption{Fraction of cooperators $f_C$ in dependence on the distribution delay $\tau$ for $r=3.6$ and inflation $q=0.99$ (see main text for details). Cooperators do not survive if $\tau \leq 3$ and fare optimally at $\tau=8$. Note also that due to inflation the overall maximum is lower and the bell narrower than in Fig.~\ref{bell} (without inflation).}
\label{inflate}
\end{figure}

Since long-term investments are frequently subject to inflation, it remains of interest to determine their impact on the described evolutionary dynamics. To do so, we assume that some fraction of accumulating public goods gets lost between two consecutive distribution events. Thus, the longer the delay, the larger the fraction that is lost. Technically this can be accommodated by multiplying the accumulated public goods with a factor $q^{\tau}$ before each distribution, where $q<1$ characterizes the rate of inflation. Although this seems to work against the success of long-term investments, we note that it may also weaken the waning moon effect because defectors will inevitably inherit a smaller amount than in the absence of inflation. Evidently, the value of $q$ should not be too small (inflation too large), as this may erase all earnings and preclude the observation of interesting collective behavior. For example, loosing 1\% ($q=0.99$) at every round sounds bearable, yet each group will have lost 40\% of the accumulated goods when reaching $\tau=50$. We thus use $q=0.99$, but note that similar behavior can be observed also when using larger inflation rates within reasonable bounds. As depicted in Fig.~\ref{inflate}, the bell-shaped dependence of $f_C$ on $\tau$ survives, and with it thus also enhanced network reciprocity and the waning moon effect. In particular, it is still beneficial to accumulate public goods for a while before distribution ($f_C=0$ if $\tau \leq 3$), although the inflation will lower the maximally attainable level of public cooperation under the same conditions, as well as narrow the window where the enhanced network reciprocity due to decelerated invasion can offset the looming waning moon effect (compare with Fig.~\ref{bell}).

In sum, we have studied the evolution of cooperation in the spatial public goods games on the square lattice, focusing in particular on revealing the impact of delayed distribution of continuously accumulating goods. Motivated by actual experience, indicating that the payoffs from investments often come with a delay, we find that intermediate delays actually can promote the evolution of public cooperation. The positive effect is due to a decelerated invasion, which strengthens network reciprocity. Interestingly, delays in distribution adversely affect the invasion speed of both strategies. Cooperators are unable to benefit from the collective investment immediately despite continuously paying the price for it, yet the defectors are also unable to exploit them continuously because of that. Defectors are ultimately those who suffer more from their decelerated invasion capabilities, as they are unable to gain the high short-term benefits that would allow them to invade in the absence of delays. As the delays grow, however, not just the accumulated wealth becomes larger, but also the probability that it will fall into the hands of defectors. We have dubbed the phenomenon the waning moon effect because of the characteristic shape of the vanishing cooperative clusters it introduces. The effect is triggered by negative curvatures of the interfaces separating cooperators and defectors, which may occur either because of random fluctuations along the invasion front or because of collision of two or more growing cooperative clusters. Defectors effectively become surrounded by cooperators that may have till then accumulated a large wealth. When distribution time comes, defectors are simply handed down the heritage from generations of cooperators, which nullifies the positive effects of enhanced network reciprocity. The larger the delays, the more likely and the more devastating the waning moon effect is going to be, until essentially all the benefits from delayed distribution vanish. We have also shown that inflation acting on the accumulating contributions narrows the interval of optimal delays, and in fact strongly jeopardizes the potential success of long-term investments. We conclude that the latter are likely to yield the desired income if ``investing into the future'' refers to the future near by.

This research was supported by the Hungarian National Research Fund (Grant K-101490) and the Slovenian Research Agency (Grant J1-4055).

\end{document}